\documentclass[preprint,preprintnumbers,amsmath,amssymb]{revtex4}
\usepackage{graphicx} 
\usepackage{dcolumn} 
\usepackage{bm} 
\usepackage{natbib}
\usepackage{hyperref}
\hypersetup{colorlinks=true,linkcolor=blue,citecolor=blue,urlcolor=blue}

\begin{document}

\title{Single Atom Impurity in a Single Molecular Transistor}

\author{S. J. Ray}
\email{ray.sjr@gmail.com}


\begin{abstract}
The influence of an impurity atom on the electrostatic behaviour of a Single Molecular Transistor (SMT) was investigated through Ab-initio calculations in a double-gated geometry. The charge stability diagram carries unique signature of the position of the impurity atom in such devices which together with the charging energy of the molecule could be utilised as an electronic  fingerprint for the detection of such impurity states in a nano-electronic device. The two gated geometry allows additional control over the electrostatics as can be seen from the total energy surfaces (for a specific charge state) which is sensitive to the positions of the impurity. These devices which are operational at room temperature can provide significant advantages over the conventional Silicon based single dopant devices functional at low temperature. The present approach could be a very powerful tool for the detection and control of individual impurity atoms in a single molecular device and for applications in future molecular electronics.
\end{abstract}



\maketitle
\setcounter{figure}{0}

\section{Introduction}

In the present day semiconductor technology, the presence of impurity atoms are essential in the  transistors for achieving enhanced conduction and better functionalities \cite{Pearson1949}. These impurity atoms which are also called as dopants are introduced randomly in these devices through the process of high energy ion implantation which places them within the channel regions of the transistors. Due to the constant reduction in the size of the transistors, the presence of the impurity atoms is becoming more and more significant in determining their performances in the nanoscale limit \cite{Daniel2011, Pierre2010}. When the effective volume of conduction in a transistor is very small, the presence of the impurity atom controls the threshold voltage ($V_{th}$) and the variability between different devices. Using the advanced single ion-implantation process \cite{Shinada2005}, it is possible to achieve precise control over the number and position of individual dopant atoms within the channel area of a MOSFET and thereby reducing the fluctuation in the $V_{th}$ between different devices. This has opened the area of `single dopant devices' where transport through the individual dopant atom controls the electrical performance of the device in the nanoscale limit \cite{Koenraad2011} which have huge potential for future applications like memory devices \cite{Hollenberg2004}, quantum computation \cite{Kane1998} etc. However, in a nano-electronic device, electrical access to these dopant atoms is only possible at very low temperatures when the single electron tunnelling through the dopant atom gives rise to a measurable current \cite{Nonnenmacher1991, Pierre2010} that limits the high temperature operation of such devices for practical realisation.

However, this limitation could be overcome by using molecular electronic devices which have the potential of operation at higher temperatures \cite{Dayen2013}. Single molecular transistor (also called as organic single electronic transistors) could be an ideal candidate for this. These single molecular devices which have the structure similar to a single electronic transistor (SET) \cite{Devort1992} are made of a large organic molecule working as the `island' or `quantum dot' placed symmetrically between two Source/Drain (S/D) electrodes and the energy levels of the molecule can be tuned by using an additional gate electrode placed in close proximity of the molecule \cite{Reed1997, Heath2003}. Electrical transport in such devices occours in an incoherent manner through a sequential tunnelling process which can be explained by the Orthodox theory of Coulomb Blockade \cite{Fulton1987}. Experimental realisation of such devices were demonstrated in a number of systems involving Benzene \cite{Reed1997}, Oligophenylenevinylene (OPV) \cite{Kubatkin2003, Osorio2007, Danilov2008}, Fullerene \cite{Park2000}, Divanadium based molecules \cite{Liang2002}, Dipyridylamide \cite{Chae2006} etc. as the active component of the device which were also supported by computational investigations to estimate the charging energies of the active molecule \cite{Kaasbjerg2008, Stokbro2010, Ray2014a} that found high degrees of agreement with the experimentally obtained results.

In this work, first principal based calculations were used within a DFT framework to investigate the performance of a double gated SMT \cite{Ray2014a} at room temperature and also in the presence of an impurity atom. For different positions of the impurity atom, the behaviour of this device was investigated to find the influence of this impurity on the charge stability diagram and compared this to the case of the device with no impurity to find the usefulness of such a system for the detection of an impurity state in a single molecular device.

\section{System description and Computational Recipe}

The single molecular transistor considered for the present investigation was made of a 1,3-Cyclobutadiene (C$_{4}$H$_{4}$) molecule working as the `dot'. This molecule was placed on top of a dielectric layer with its atomic plane lying parallel to it as illustrated in Fig.~\ref{fig.1}(a) in a double gated geometry as also reported in \cite{Ray2014a}. The Source/Drain (S/D) electrodes were placed symmetrically around the molecule with a minimum separation $\sim$ 4 \AA\, from the nearest Hydrogen atoms from each electrodes. The `top' and `bottom' gates ($\sim$ 1 \AA\, thickness) were placed accordingly backed by the dielectric layers (with a dielectric constant of $10\varepsilon_{0}$) which are of thicknesses $d_{t}$ = 2 \AA\, and $d_{b}$ = 3.7 \AA\, respectively. The separation between the molecule and the top and bottom dielectric layers were 5 \AA\, and 1.5 \AA\, respectively. A single Silicon atom was introduced in this SMT as an impurity along the central $yz$-plane in 3 separate positions on top of the molecule as illustrated by the `red' circle in Fig.~\ref{fig.2}(b),(c),(d). These positions (and the devices) were labelled following the horizontal distance of the impurity from the left side of the `Source' electrode (i.e. the $xy$-plane at $z=0$) and labelled as $z6$ [for impurity placed at a distance 6 \AA\, from the $z=0$ plane in Fig.~\ref{fig.2}(b)], $z9.5$ [for impurity placed at a distance 9.5 \AA\, from the $z=0$ plane in Fig.~\ref{fig.2}(c)], $z12$ [for impurity placed at a distance 12 \AA\, from the $z=0$ plane in Fig.~\ref{fig.2}(d)] and $z0$ for the case of no impurity [Fig.~\ref{fig.2}(a)]. For the $z6$, $z9.5$ and $z12$ cases, the $x$ and the $y$ coordinates  were fixed at 6.88 \AA\, and 8 \AA\, respectively from the (0, 0, 0) corner. In the $z9.5$ position, the Si atom was placed immediately on top of the centre of the molecule while for the $z6$ and $z12$ cases, the Si atom was placed at a distance 3.5 \AA\, from the left and 2.5 \AA\, to the right of the Si atom in $z9.5$ position (along the $z$-axis) respectively. Gold was used as the generic electrode material for a minimum contact resistance with a work function ($W$) of 5.68 eV.

Ab-initio calculations were performed to estimate the charging energy of the molecule in a SMT,  using the Atomistic ToolKit Package (ATK) developed by the Quantum Wise Division \cite{ATK}. This  uses the recipe introduced by Stokbro $et.~al$ \cite{Stokbro2010} based on the formalism proposed by Kaasbjerg $et.~al$ \cite{Kaasbjerg2008}. This method uses a nonspin-polarised DFT framework for expanding the wave functions in a double-$\zeta$ polarised basis set under the Local Density Approximation (LDA). Details of the computational procedure can be found here \cite{Stokbro2010}.

\section{Results and Discussion}

\begin{figure*}
\begin{center}
\includegraphics[width=17cm]{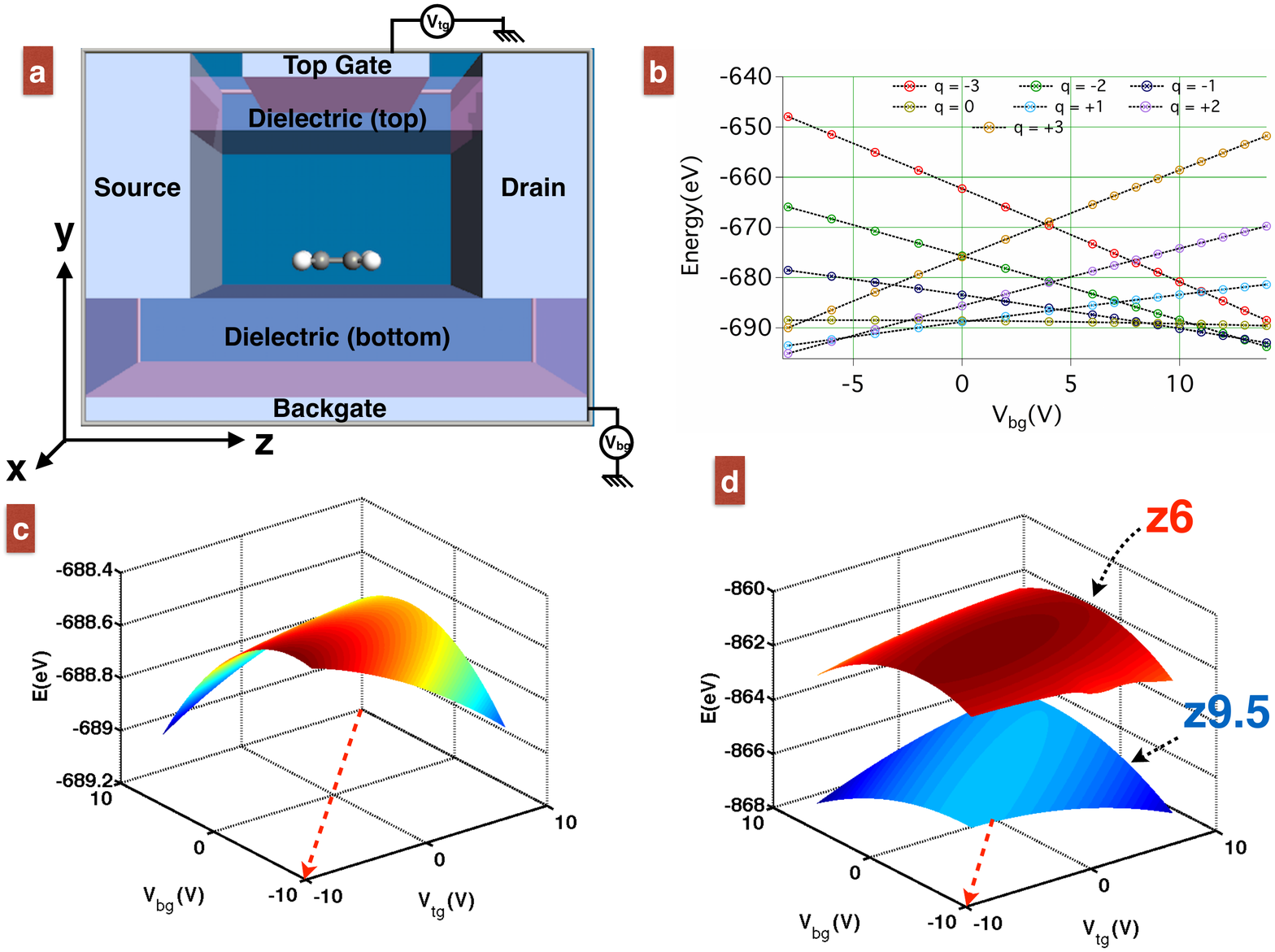}
\caption{{\small (a) Schematic of the SMT device used in the present work in the absence of any impurity. The 1,3-Cyclobutadiene molecule stays at the centre of the device with its atomic plane lying parallel to the bottom dielectric layer equidistant from the S/D electrodes. (b) Total energy of the molecule for different charge states as function of $V_{bg}$ for a fixed $V_{tg}$ = -8V for $z0$ device. The `points' are the data and the dotted lines are the fitted curves made using Eqn.~\ref{eqn.1}. (c) Total energy of the molecule as functions of the two gate voltages for the charge state $q = 0$ in the absence of any impurity atom ($z0$) and (d) Energy surface for the $q = 0$ state for $z6$ and $z9.5$ positions in the presence of impurity in the device.}}
\label{fig.1}
\end{center}
\end{figure*}

For a given charge state, the total energy of the molecule for the $z0$ SMT was plotted as function of the backgate voltage ($V_{bg}$) for different charge states in Fig.~\ref{fig.1}(b) for a fixed top gate voltage ($V_{tg}$) = -8V. At $V_{bg} < 0$, the energy for the positive charge states are smaller than the energy for the negative charge states due to energy stabilisation which is opposite for these charge states at $V_{bg} >0$.  For a given charge state, the energy at a specific $V_{bg}$ can be described analytically by,
\begin{equation}
E(q, V_{bg}) = E_{0}(q) + \alpha qV_{bg} + \beta (eV_{bg})^{2}
\label{eqn.1}
\end{equation}
The linear (2$^{nd}$) term in Eqn.~\ref{eqn.1} represents the direct coupling between the molecule and the gate electrode and the value of $\alpha$ represents the strength of coupling which for the present case was estimated to be $\sim$ 0.59. The $3^{rd}$ term is independent of the charge state and signifies the contribution of electrical polarisation under the influence of an electric field. The value of $\beta \sim$ 0.002 eV$^{-1}$ and the fact that $\beta < \alpha$ indicate the dominance of direct molecule-gate coupling for the present device.

For the case of the SMT without any impurity ($z0$), the energy surface for the neutral state ($q = 0$) of the molecule was plotted as functions of the gate voltages in Fig.~\ref{fig.1}(c). In the absence of a non-zero charge state, the main contribution towards the total energy was contributed by the electrical  polarisation of the molecule and thus symmetrically curved around the diagonal (red line in Fig.~\ref{fig.1}(c)) on the voltage plane. Comparing this to the case of a SMT with an impurity when the molecule in the similar charge state, the energy surfaces were found to be completely different as illustrated in Fig.~\ref{fig.1}(d) for the cases of $z6$ and $z9.5$ devices. Firstly making a systematic comparison between the different energy surfaces, one can see that under identical device geometry and electrical conditions, the energy of the molecule gets significantly reduced under the introduction of the impurity. The maximum in energy for the $z0$ case in Fig.~\ref{fig.1}(c) is $\sim$ -688.6 eV, while the maximum in the $z6$ and $z9.5$ cases ranges between -862 eV and -867 eV. In the neutral state of the molecule, the impurity atom screens additional polarisation to the molecule and this leads to a reduction of the total energy. The difference of the energy between the $z6$ and $z9.5$ cases is relatively smaller compared to the $z0$ case which occurs as the the movement of the impurity atom between the $z6$ and $z9.5$ cases brings in a relatively smaller change in the induced polarisation compared to situation when the impurity was completely absent in the device. Nevertheless, an observable difference in the two energy surfaces can be found in Fig.~\ref{fig.1}(d) for two different positions of the impurity within the device. Under the presence of a top gate, in the $z6$ case the distance between the impurity and the molecule is not identical for all the atoms and the difference in the electric fields experienced by different atoms is larger compared to the $z9.5$ case, when the impurity is positioned symmetric to all the atoms of the molecule. Hence, the total energy of the molecule is smaller for the $z9.5$ case when the impurity stays directly on top compared to the $z6$ situation when the influence of electrical polarisation due to the top gate is larger on the molecule.

\begin{figure*}
\begin{center}
\includegraphics[width=17cm]{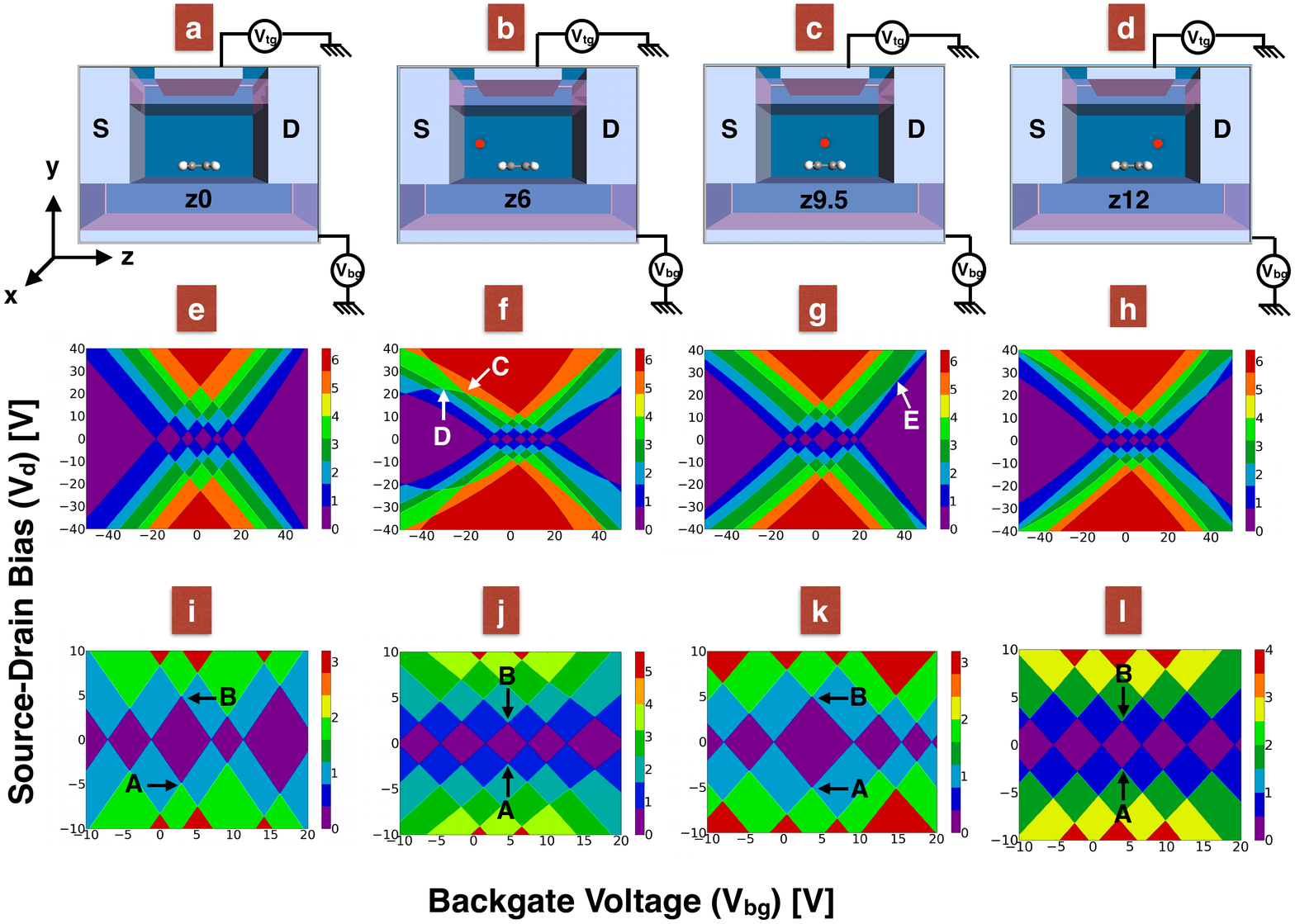}
\caption{{\small [Top panel (a)-(d)] Schematic of the SMT devices for different positions of the Si atom (in red) (details mentioned in the text): (a) In the absence of impurity ($z0$ position), (b) Close to the the Source electrode ($z6$ position), (c) Centred on the device immediately on top of the molecule ($z9.5$ position) and (d) Close to the `Drain' electrode ($z12$ position). [Central panel (e)-(h)]  The charge stability diagram at $V_{tg}$ = 0 V for different positions of the impurity atom as illustrated in the respective top panels i.e (e) in $z0$ position, (f) $z6$ position, (g) $z9.5$ position, (h) $z12$ position, [Bottom panel (i)-(l)] : Zoomed version of central regions of the charge stability diagrams used for investigating the influence of the impurity as illustrated in the respective central panel, (i) $z0$ position, (j) $z6$ position, (k) $z9.5$ position, (l) $z12$ position. The points `A' and `B' refer to the coordinates of the two vertices of the central diamond in the ground state of the molecule.}}
\label{fig.2}
\end{center}
\end{figure*}

Details of the impurity detection procedure can be understood from the charge stability diagram illustrated in Fig.~\ref{fig.2}. Due to the sequential nature of the transport arising from the weak coupling between the molecule and the S/D electrodes, the electronic transport is only possible when the charging energy of the molecule ($E_{ch}$) satisfies the condition :  $e|V_{d}|/2 > (E_{ch} + W)> -e|V_{d}|/2$ for a finite source-drain bias ($V_{d}$). This results in the charge stability diagram in the form of diamond shaped regions on the voltage plane that indicates the separation between the conducting and non-conduction regions of transports for different values of $V_{d}$ and $V_{g}$. In Fig.~\ref{fig.2}, the stability diagrams were plotted for different positions of the impurity for a fixed value of $V_{tg}$ = 0 V. Just for clarity, it is to be noted that the images in each column are for the same position of the impurity in the device. From naked eyes, differences in the stability diagrams can be found following the colour scales for different positions of the impurity. In the absence of any impurity ($z0$), the region of crossings between different charge states are almost linear as illustrated in Fig.~\ref{fig.2}(e) which changes significantly when the impurity is introduced in the device as illustrated in Fig~\ref{fig.2}(f-h).  The boundary between the $q = 3$ and $q = 4$ states for $V_{bg} < 0$ is curved which continues to be of similar nature for the boundary between other charge states for further negative values of the $V_{bg}$ as can be seen following the line joining the points `C' and `D' in Fig.~\ref{fig.2}(f). Moving to the case of $z9.5$ in Fig.~\ref{fig.2}(g), when the impurity is placed along the centre of the molecule, some curved boundaries could be seen on the stability diagram near point `E' for $V_{bg}>0$ which indicates the crossing between $q = 1, 2, 3$ states. In Fig.~\ref{fig.2}(h) for the $z12$ position when the impurity lies closer to the drain electrode, the boundary between the $q = 0, 1, 2$ states are also found to be curved at $V_{bg} > 0$. As the S/D electrodes were considered to be identical in the present analysis (unlike the case of a real experimental situation where minor differences exists between them), hence the charge stability diagram will stay the same for an exchange of the S/D electrode positions due to the symmetry reasons.

\begin{figure*}
\begin{center}
\includegraphics[width=11cm]{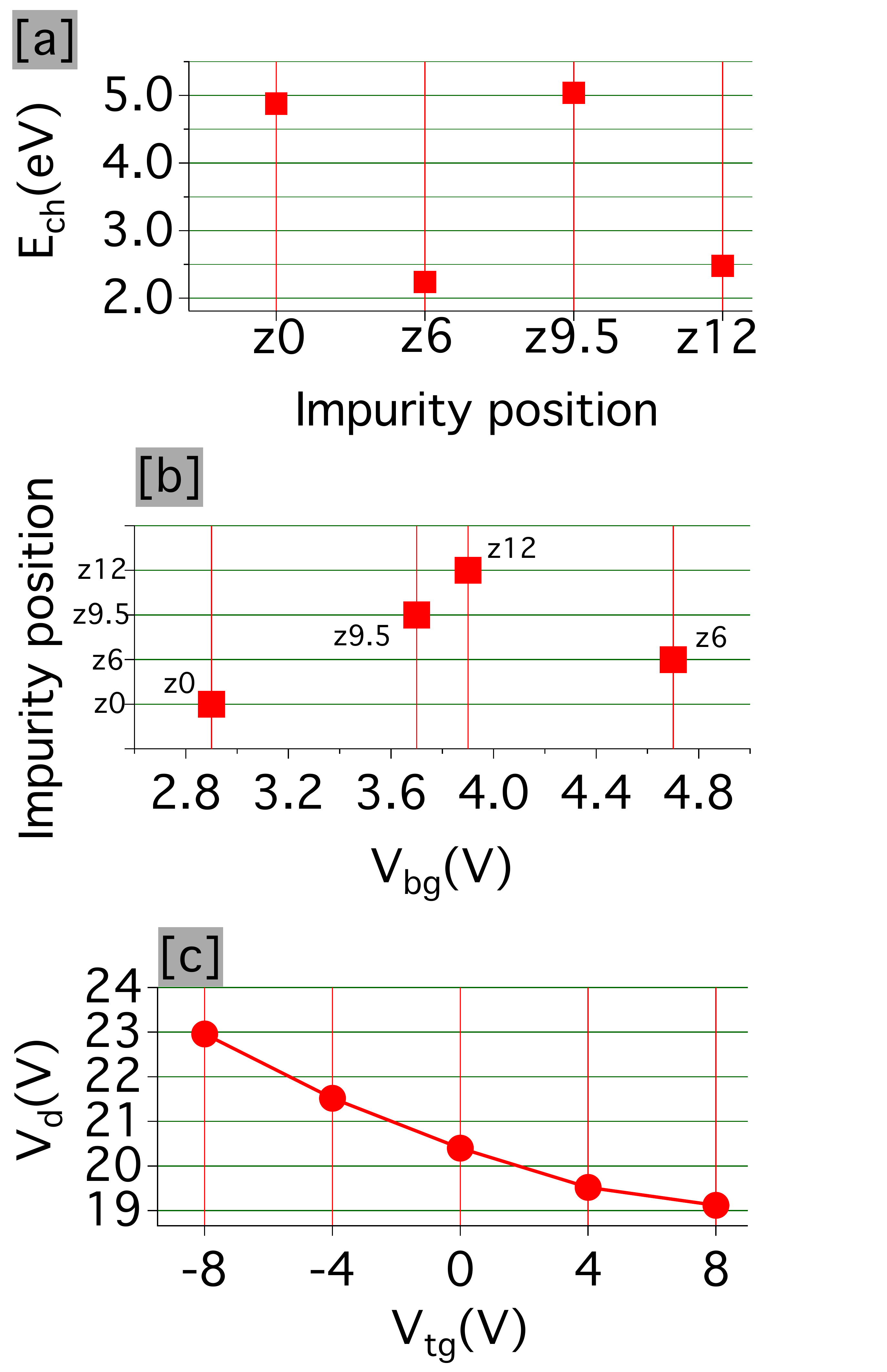}
\caption{{\small (a) Charging energy for the $q = 0$ case of the molecule in the ground state as estimated from the height of the central diamond illustrated in Fig.~\ref{fig.2} for different impurity positions, (b) Relative position of the bottom vertex (A) of the central diamond estimated from the value of $V_{bg}$ from the bottom panel of Fig.~\ref{fig.2}. (c) $V_{d}$ value as function of $V_{tg}$ as estimated from point `C' in Fig.~\ref{fig.2}(f).}}
\label{fig.3}
\end{center}
\end{figure*}

In order to make further analysis of the influence of the impurity, systematic comparisons were made between the charge stability diagrams focussing on the central regions of the Coulomb diamonds. In Fig.~\ref{fig.2}(i),(j),(k),(l), such regions were illustrated as taken from the Fig.~\ref{fig.2}(e),(f),(g),(h) respectively. The bottom vertex of the central diamond `A' was considered as a reference point where 3 charge state $q = 0,1,2$ intersect as illustrated in Fig.~\ref{fig.2}(i) and the evolution of this point for different positions of the impurity was analysed to speed up the detection process. As can be seen in Fig.~\ref{fig.2}(i - k), the coordinate of the point `A' changes for different positions of the impurity atom and the value of the $V_{bg}$ ($x$-coordinate of `A') was plotted for these cases in Fig.~\ref{fig.3}(b).
It can be said that the for different positions of the impurity, the value of the $V_{bg}$ is different and significantly distant from the $z0$ value, in the absence of an impurity. There is no direct overlap of $V_{bg}$ between them and it is easy to identify different position of the impurity using this. This value of $V_{bg}$ obtained from such a reference point could be used for tracking the location of the impurity within the SMT. In Fig.~\ref{fig.3}(b), the difference in $V_{bg}$ between the $z0$ and $z6$ is much larger compared to the differences between the $z0\leftrightarrow z9.5$ or $z0\leftrightarrow z12$. In the $z6$ case, the impurity stays farthest from the molecule compared to the $z9.5$ and $z12$ cases which indicates that maximum sensitivity for detection will occour when the impurity is placed close to the S/D electrodes. In the $z6$ case, the impurity is more decoupled to the molecule compared to the other two cases and due to its proximity to the S electrode, any transport will be significantly influenced by the impurity due to its location.

Another way of looking for the evidence of impurity was performed via comparing the charging energy in different cases. The charging energy ($E_{ch}$) was estimated for the ground state of the molecule in the neutral case from the height of the central diamond from Fig.~\ref{fig.2}(i-l). In Fig.~\ref{fig.3}(a), $E_{ch}$ was plotted for different positions of the impurity atom which also changes systematically with the impurity position and different from the value in the no-impurity situation. The charging energy in the $z0$ and $z9.5$ cases are much higher compared to the $z6$ and $z12$ cases. For the $z6$ case, the impurity stays at a corner of the device and only the H-atoms of the molecule close to it sense maximum influence from it. However, when the impurity stays on top along the centre of the molecule in the $z9.5$ position, the overlap of the impurity wave function with the molecular orbitals is significantly higher that gives rise to a higher charging energy in the neutral state. The similarity of the value of charging energy in the $z6$ and $z12$ can be explained due to their almost symmetric distances from the molecule. The sensitivity of the charging energy to the impurity position can also be considered as a fingerprint to identify the position of the impurity in such a device. It is to be noted that similar changes in value of the $V_{bg}$ and $E_{ch}$ were also observed when the impurity position was varied along the $x$-axis keeping the ($y,z$) coordinates fixed which can also be systematically realised from the charge stability diagrams and the energy surfaces. Due to symmetry reasons, for varied position along the $y$-axis keeping $x,z$ coordinates fixed, influence of the impurity position was present on the charge stability diagrams \cite{SI}. Thus this technique which carries unique information about the location of the impurity in all 3 geometrical directions can be used as a very powerful tool for the detection of such impurity state in a single molecular device.

In an experimental situation, usually the impurity atom was placed via controlled ion implantation process within the channel region of a CMOS device and the effect of the impurity can only be observed at low temperature when the thermal excitations are very low. However, in the present case, the organic molecule based SMT is found to be operational at 300K (room temperature) and the effect of impurity is significantly prominent in all the cases that can be realised directly from the charge stability diagram, charging energy behaviour or the dependence of total energy with the gate voltages. Unlike the discrete charge states, experimentally the charge stability diagram is constructed by using the conductance or the differential conductance ($dI/dV$) on the voltage plane. If the impurity stays in the 'OFF' region of the device, then its presence can be found as a discrete peak in the $I(V_{g})$ scan and if the impurity is coupled to the two gates as described in the present case, then on a 2D conduction map on the gate voltage planes, it can be realised in the form of a `triple point' depending of the strength of the impurity-dot coupling.

It is also the noted that in a double gated geometry, the top gate electrode allows additional operational control over the electrostatics of the system. On the charge stability diagram, the central diamond occours in a $V_{bg}$ range which is different for different cases and this can be shifted by controlling the $V_{tg}$ due to almost linear nature of the gate-dot coupling. This effect was realised by considering the systematic evaluation of the point `C' with $V_{tg}$ [taken from Fig.~\ref{fig.2}(f)] and the $V_{d}$ values estimated from the coordinate of `C' was plotted as function $V_{tg}$ in Fig.~\ref{fig.3}(c). The change of $V_{d}$ with the change of $V_{tg}$ is almost linear under identical electrostatic conditions and this shows that it is possible to find the desired region of operation by tuning the value of $V_{tg}$ on the charge stability diagram.

\section{Conclusion}

To summarise, using First-principle based calculations in the Coulomb Blockade regime, the electrostatic behaviour of a double gated SMT was investigated to understand the presence of an impurity atom within it. Due to the extreme sensitivity of the molecular energy levels to the gate excitation in the weak-coupling limit, the presence of the impurity atom can be realised from the charge stability diagram which is distinctive for different positions of the impurity atom. Focussing around a specific region of the charge stability diagram, it is possible to uniquely identify the signature of the location of the impurity atom in such a device. The charging energy of the molecule has been found to be sensitive to the position of the impurity atom that could also be used as an identification technique. For a specific charge state of the molecule, the influence of the impurity atom can also be found from the energy surfaces (as functions of the gate voltages) which changes with the position of the impurity atom in the SMT. These together with the charge stability diagram could be used as an electronic fingerprint for the detection of the presence and location of an impurity atom in a nanoscale device. The room temperature operational feasibility of such single molecular systems allows the possibility of the use of such devices in future molecular electronics as an alternative to the Si-based single dopant devices where operational feasibility is limited by low operational temperatures.


\bibliographystyle{apsrev}


\newpage 

\end{document}